# Coalescence of Multiple Topological Orders in Quasi-One-Dimensional Bismuth Halide Chains

Jingyuan Zhong[1,#], Ming Yang[1,2,#], Wenxuan Zhao[3], Kaiyi Zhai[3], Xuan Zhen[4], Lifu Zhang[5], Dan Mu[6], Yundan Liu[6], Zhijian Shi[1], Ningyan Cheng[7], Wei Zhou[8], Jianfeng Wang[1], Weichang Hao[1], Zhenpeng Hu[4,*], Jincheng Zhuang[1,*], Jinhu Lü[9], Yi Du[1,*]

[1] School of Physics, Beihang University, Haidian District, Beijing 100191, China

[2] The Analysis & Testing Center, Beihang University, Beijing 102206, China.

[3] State Key Laboratory of Low Dimensional Quantum Physics, Department of Physics, Tsinghua University, Beijing 100084, China

[4] School of Physics, Nankai University, Tianjin, 300071, China

[5] School of Biomedical Engineering and Technology, Tianjin Medical University, Tianjin 300070, China

[6] Hunan Key Laboratory of Micro-Nano Energy Materials and Devices, and School of Physics and Optoelectronics, Xiangtan University, Hunan 411105, China

[7] Information Materials and Intelligent Sensing Laboratory of Anhui Province, Key Laboratory of Structure and Functional Regulation of Hybrid Materials of Ministry of Education, Institutes of Physical Science and Information Technology, Anhui University, Hefei 230601, China

[8] School of Electronic and Information Engineering, Changshu Institute of Technology, Changshu 215500, China

[9] School of Automation Science and Electrical Engineering, Beihang University, Beijing 100191, China

#Jingyuan Zhong and Ming Yang contributed equally to this work.

*Correspondence authors. E-mail: jincheng@buaa.edu.cn; zphu@nankai.edu.cn; yi_du@buaa.edu.cn

**Abstract:**

**Topology is being widely adopted to understand and to categorize quantum matter in modern physics. The nexus of topology orders, which engenders distinct quantum phases with benefits to both fundamental research and practical applications for future quantum devices, can be driven by topological phase transition through modulating intrinsic or extrinsic ordering parameters. The conjoined topology, however, is still elusive in experiments due to the lack of suitable material platforms. Here we use scanning tunneling microscopy, angle-resolved photoemission spectroscopy, and theoretical calculations to investigate the doping-driven band structure evolution of a quasi-one-dimensional material system, bismuth halide, which contains rare multiple band inversions in two time-reversal-invariant momenta. According to the unique bulk-boundary correspondence in topological matter, we unveil a composite topological phase, the coexistence of a strong topological phase and a high-order topological phase, evoked by the band inversion associated with topological phase transition in this system. Moreover, we reveal multiple-stage topological phase transitions by varying the halide element ratio: from high-order topology to weak topology, the unusual dual topology, and trivial/weak topology subsequently. Our results not only realize an ideal material platform with composite topology, but also provide an insightful pathway to establish abundant topological phases in the framework of band inversion theory.**

## Introduction:

Searching for yet topological phases has garnered immense interest, not only in the research community but also for low-energy-low-cost technology and quantum information science due to the existence of topologically protected conduction channels. The topological band theory proposed by Kane and Mele uses the concept of the topological invariant to classify the topological quantum materials with robust boundary states protected by time-reversal-symmetry [1-4], where the topological phases are achieved by band inversion with varied eigenvalues of wavefunction parities of the occupied states after turning on the spin-orbit coupling (SOC). The prototypical examples of these materials, the strong (weak) topological insulators, can be realized by such band inversion occurring at odd (even) numbers of time-reversal-invariant (TRI) momenta, which has been experimentally confirmed in recent decades under the condition that the band inversion only involves one conduction band and one valence band at each TRI momentum [5-14]. The situation becomes complicated when multiple band inversions (MBI) take place in TRI momentum, leading to versatile topological phases due to the additional freedom in the modulation of band inversion. In fact, the MBI has been investigated under the condition of various characteristics [15-24]. For instance, the multiple Dirac cones are derived from the combined effect of Rashba splitting and MBI, which have the potential to unveil the intertwined physical phase by Rashba spin physics and topological nontriviality [15-17, 23]. The strain effect can realize the MBI and the consequent weak topological insulating phase or semi-Dirac semimetal phase through the topological phase transition (TPT) [19-20]. The topological compensated semimetal phase and topological excitonic insulating phase are identified in the MBI systems under the strong SOC strength [17, 22]. Moreover, the quasi-one-dimensionality is benefit to the emerge of MBI at different TRI momenta, dedicating to the unusual topological phase such as high-order topological insulator (HOTI) [21]. Therefore, searching for the appropriate materials with multiple conduction band and valence band pairs involved in the SOC-induced band inversions can extend the experimental study of the unusual topological phases correlated with the topological band theory.

## Results:

Three-dimensional (3D) $α'$-Bi$_4$Br$_4$, which is constructed from one-dimensional (1D) chains, features double band inversions around Fermi level at two TRI momenta, at the $M$ point and the $L$ point of the 3D Brillouin zone (BZ), as displayed in Fig. 1**a-c**. For the $L$ point, the opposite signs (+−) of the parity eigenvalues near the Fermi level for the bulk valence bands (BVBs) are retained after the inclusion of SOC, leading to a trivial state ($\mathbb{Z}_2 = 0$). In contrast, the parity eigenvalues of BVBs vary from (−−) to (++) at the $M$ point, resulting in a HOTI state ($\mathbb{Z}_4 = 2$) with gapped surface states but 1D helical gapless hinge states [21, 25-27]. Thus, $α'$-Bi$_4$Br$_4$ is an excellent parent phase to use for designating the special topological phases by varying the eigenvalues at different TRI momenta through importing perturbations. Leveraging the combined techniques of angle-resolved photoemission spectroscopy (ARPES) and scanning transmission microscopy (STEM), we identify the unusual dual topology modes, strong topological insulator (STI) and HOTI, which simultaneously coexist in this material system after applying the isocharge doping of I atoms on Br sites. Moreover, the interplane stacking orders are also modulated varying the ratio between the halide elements, engendering abundant topological phases in this quasi-1D system. Accordingly, a topological phase diagram of Bi$_4$(Br$_{1-x}$I$_x$) as a function of I concentration was established to provide a route to precise control of a single crystal quantum system.

A structural model of $\alpha'$-Bi$_4$X$_4$ with the $C$2m symmetry group is shown in Fig. 1**a**, where the structural building block is the quasi-1D bismuth chain passivated by halide atoms, which are connected with each other through the van der Waals (vdW) force along two directions (the lattice $a$ direction and the lattice $c$ direction). Therefore, there are two cleavable planes, the (001) surface and the (100) surface, which were suitable for the surface characterizations in this system. Fig. 1**d** and 1**e** displays the scanning tunneling microscope (STM) topographies of the cleaved (001) and (100) surfaces of Bi$_4$Br$_4$, respectively, where large flat planes and lattice constants can be identified. High-angle annular dark-field – scanning transmission electron microscopy (HAADF-STEM) combined with the focused ion beam (FIB) technique was applied to characterize the structure of the uncleavable (010) plane. The antiparallel nature of two adjacent layers is revealed in Fig. 1**f**, forming the A-B stacking mode and consistent with the cross-sectional view of the structural model in Fig. 1**a** as well as previous reports [27-28]. The high quality of these single crystals is indicated by the sharp diffraction peaks in X-ray diffraction (XRD) measurements (see Supplementary Fig. 1 for details). For the single crystals doped with I atoms, the actual I concentration was measured by the energy-dispersive spectroscopy (EDS), which shows a monotonic change with the nominal I concentration in the raw materials (see Supplementary Fig. 2 for details). We refer to the actual I concertation obtained from the EDS measurements below as the I concentrations. There is a redshift of the positions of (00$l$) peaks in the XRD spectra with increasing I concentrations, implying the possible increase of the $c$-axis lattice constant. HAADF-STEM measurements on the (010) planes of the different samples were performed to determine the positions of the I atoms and their influence on the stacking arrangement along the lattice $c$ axis direction. The high-resolution STEM images in Fig. 1**f** show that there are no atoms in the intercalation positions, indicating that the Br atoms were successfully replaced by I atoms. Moreover, most of the samples possess the same A-B stacking mode ($\alpha'$ phase) expected for the samples with I concentrations of 0.5 and 1, in line with the results on powder samples in a previous work [29]. We categorize the structures of Bi$_4$(Br$_{0.5}$I$_{0.5}$)$_4$ and Bi$_4$I$_4$ as $\gamma$ phase and $\alpha$ phase, respectively, and the corresponding STM topography of the (001) planes with crystal parameters can be seen in Supplementary Fig. 3. Figure 1**g** displays the lattice constants extracted from our XRD curves, STM images, STEM results, and other references [21, 25, 29-33] as functions of I concentration, where all three values increase with increasing I content, which is due to the larger ionic radius of I compared to Br. It should be noted that the increased amplitudes of the lattice constants for $a$ axis and $c$ axis are much larger than that of $b$ axis, implying that the I doping mainly affects the interchain space rather than the intrachain space. All these results demonstrate that the lattice constants of Bi$_4$Br$_4$ are modulated by I doping and that the $\alpha'$ phase can be retained over a large concentration range.

We first performed ARPES measurements on the *in-situ* cleaved (001) surfaces of Bi$_4$(Br$_{1-x}$I$_x$)$_4$ crystals to reveal the bulk-structure evolution with the I content due to the fact that there is no surface state in the (001) plane [34]. 3D plots of the ARPES spectra on the (001) surface in reciprocal ($k_x$, $k_y$, $E$) space, where $k_x$ and $k_y$ are wave vectors and $E$ is the energy, for samples with different I content are shown in Fig. 2**a**. An elliptical electron pocket emerges at the $\overline{\text{M}}$ point of the (001) surface BZ, and no band can be observed at the $\overline{\Gamma}$ point of the (001) surface BZ near the Fermi level (see Supplementary Fig. 4 for details). These ARPES spectra are well consistent with the calculation results in Fig. 1**c**, as both the $M$ point and $L$ point of the 3D BZ, where the near-Fermi-level bands are located, are projected into the $\overline{\text{M}}$ point for the two-dimensional (2D) BZ of the (001) plane. Figure 2**b** shows the energy-momentum dispersion at the $\bar{\text{M}}$ point cut along the chain direction ($k_y$) of the samples with different I content. A bulk energy gap $\Delta_B$ can be observed, and the gap size initially decreases, reaching a minimum value, and then increases with the increasing I content. The value of $\Delta_B$, which can be estimated through energy differentiation between the edges of the full width at half maximum of the top valence bands and bottom conduction bands in the energy distribution curves (EDCs), as displayed in Fig. 2**c**, is 135 meV, 77 meV, 48 meV, 17 meV, 29 meV, and 137 meV for the crystals with I contents of 0%, 25%, 50%, 62%, 80%, and 100%, respectively (see Supplementary Fig. 5 for details). This nonmonotonic trend in the gap value implies possible gap closing and band inversion at the critical point near 70% I concentration. In fact, our temperature-dependent resistance measurements of different I content samples show that the 73% I concentration sample is the only one with metallic behavior (see Supplementary Fig. 6 for details), which indicates that the bulk gap is possibly closed and agrees well with the gap size trend in the APRES spectra. Additionally, the density functional theory (DFT)-calculated band structures of 3D $Bi_4(Br_{1-x}I_x)_4$ show a band inversion at $x = 0.75$ (see Supplementary Fig. 7 for details), in strong agreement with the experimental observations.

Since both the *M* point and the *L* point of the 3D BZ are projected into the $\bar{\text{M}}$ point of the 2D BZ of the (001) plane in which the band inversion occurs, it is hard to distinguish the evolution of the 3D band structure between these two points from the ARPES spectra of the (001) plane. Nevertheless, the *M* point and the *L* point of the 3D BZ are separately projected into the $\bar{\Gamma}$ point and the $\bar{\text{Z}}$ point of the 2D BZ of the (100) plane, as indicated in Fig. 1**b**. Thus, the (100) surface characterizations are expected to provide more information on the band structure. The cleaved (100) surface is rough compared to the cleaved (001) surface due to the stronger interchain coupling strength along the lattice *a* direction than along the lattice *c* direction [35-36]. We applied the laser-based ARPES measurements with micrometre light size to characterize the electronic structure of the *in-situ* cleaved (100) surface of these single crystals to acquire high-resolution spectra. Unlike the (001) surface, the ARPES spectra of the (100) surface exhibit surface states (see Supplementary Fig. 8 for details). 3D plots of ARPES spectra on (100) surface in ($k_y$, $k_z$, $E$) space for different crystals are shown in Fig. 3**a**, where the highly anisotropic electronic structures of 2D surface states are identified because of the 1D nature of their building blocks. The $E$-$k_y$ dispersion spectra of different crystals at the $\bar{\Gamma}$ point and $\bar{\text{Z}}$ point are presented in Fig. 3**b** and 3**c**, respectively. For $\alpha'$-$Bi_4Br_4$, the surface states at both the $\bar{\Gamma}$ point and $\bar{\text{Z}}$ point are constructed from two nondegenerate Dirac bands with surface energy gaps $\Delta_S$ that have opened at the band crossing points away from the TRI momenta. The values of $\Delta_S$ for $Bi_4Br_4$ are ~38 meV and ~35 meV around the $\bar{\Gamma}$ point and $\bar{\text{Z}}$ point, respectively (see Supplementary Fig. 9-10 for details of the gap analysis). After introducing the doped I atoms, the surface bands of the (100) plane show subtle changes, such as varied Fermi velocities and decreased $\Delta_S$ at both the $\bar{\Gamma}$ point and the $\bar{\text{Z}}$ point, in the condition of the fixed $\alpha'$ phase. Interestingly, there is transition from the double Dirac band structure to the single Dirac band structure at the $\bar{\text{Z}}$ point of the (100) BZ of $\alpha'$-$Bi_4(Br_{0.2}I_{0.8})_4$, which has been confirmed by the second derivative results in Supplementary Fig. 11. It is notable that $Bi_4(Br_{0.2}I_{0.8})_4$ belongs to a different phase from the HOTI region, as its I content (0.8) is just a step over the critical point (~0.73) (Fig. 2**b**). Hence, the spectra of the (100) surface states coincide with the bulk structures obtained in ARPES results measured in the (001) plane, confirming the occurrence of a phase transition.

The (001) monolayer $Bi_4Br_4$ has been predicted and subsequently confirmed by experiment to be a large gap (~ 0.2 eV) quantum spin Hall insulator (QSHI) [25,34-35]. The 3D $α'$-Bi$_4$Br$_4$ can also be regarded as an alternating arrangement of two antiparallel (001) monolayers with only in-plane inversion symmetry (see Supplementary Fig. 12 for details). Our first-principle calculations of the band structure of two adjacent layers (A layer and B layer) in Supplementary Fig. 12 imply nondegenerate electronic structures, which has been experimentally observed from scanning tunneling spectroscopy curves (see Supplementary Fig. 13 for details). The structural discrepancy lifts the degeneracy of two neighboring (001) monolayers (see Supplementary TABLE 3 for details), and the two pairs of BVBs and bulk conduction bands (BCBs) at either the $M$ point or the $L$ point of the 3D BZ near the Fermi level in Fig. 1c are contributed by the A layer and the B layer, respectively. Therefore, the 1D edge states of the two QSHI layers, the A layer and the B layer, are also nondegenerate with the different Fermi velocities and energy positions of the Dirac point. These nondegenerate edge states are arranged along the lattice $c$ direction to form the (100) surface states with two Dirac band structures and a surface gap opened at their crossing points by the finite interlayer interaction (see Supplementary Fig. 12 for details). It should be noted that multiple Dirac bands have been resolved in the MBI system with different surface terminations, which supports the observation of two sets of Dirac bands in $α'$-Bi$_4$Br$_4$ since both of the two edges are exposed to form the (100) plane [15-17]. Consequently, these two nondegenerate Dirac bands can be directly treated as the result of the double band inversions at both the $M$ point and the $L$ point of the 3D BZ under the action of SOC, and the single Dirac band at the $\bar{Z}$ point of the 2D (100) surface BZ of Bi$_4$(Br$_{0.2}$I$_{0.8}$)$_4$ is derived from the transition from double band inversion to single band inversion at the $L$ point of the 3D BZ, which is confirmed by the surface band calculations of (100) plane of the heavily I doped crystal (see Supplementary Fig. 7 for details).

The HOTI nature of $α'$-Bi$_4$Br$_4$ is correlated with the change of the parity eigenvalues from (——) to (++) at the $M$ point of the 3D BZ after switching on SOC. The double band inversion at the $L$ point makes no contribution to the symmetry-indicator topological nontrivial invariants due to the invariable parity eigenvalues (—+), as displayed in Fig. 4a and 4b. The transition from double band inversion to the single band inversion at the $L$ point engenders the typical topological transition to the STI as the only inverted parity eigenvalue, as shown in Fig. 4c. Additionally, the double band inversion at the $M$ point responsible for the HOTI characteristic is unchanged, giving rise to the unusual dual topological phase of HOTI and STI in $α'$-Bi$_4$(Br$_{0.2}$I$_{0.8}$)$_4$.

According to the above results, several TPTs are revealed in this quasi-1D bismuth halide system, and the electronic phase diagram of Bi$_4$(Br$_{1-x}$I$_x$) is plotted in Fig. 4d. $α'$-Bi$_4$Br$_4$ is treated as the parent phase in this system. The $Δ_S$ resulting from the interlayer interaction is correlated with the HOTI gap, where the 1D hinge states reside. With increasing I concentration, both the $Δ_B$ and $Δ_S$ are diminished, which may be caused by the enlarged interchain space identified by our structural results in Fig. 1 and the reduced interchain coupling. Thus, the strain effect can be the main reason for the reduction of gap values (See Supplementary Fig. 15). With further increased I content, $Δ_B$ continually decreases and reaches zero at the critical point of I content around 0.73 (labelled by the green ball). After this TPT, the bulk gap reopens and the double band inversions at the $\bar{Z}$ point turn into a single band inversion. The composite topological phase, STI plus HOTI, is generated with the coexistence of 2D surface states and 1D hinge states. The existence of the hinge states in this composite topological phase is supported by the calculation results of 20*$a$ × 20*$c$ $α'$-Bi$_4$(Br$_{0.25}$I$_{0.75}$)$_4$ model (see Supplementary Note 6 for details). Although these band inversions should induce a $Z_4 = 3$ state and STI order according to the original theory [37], the recent work indicates that the similar composite topological order can be realized in elemental As film [38]. It should be noted that the crystal structures are altered when the I content are 50% and 100%, forming the γ phase and α (β) phase, respectively. The γ phase is a structure including three alternately arranged layers, corresponding to a non-degenerate weak topological insulator (WTI), while the α phase and β phase are ascribed to a topological trivial insulator and a WTI according to topological band theory, respectively [39]. It should be noted that two Dirac bands similar to α'-Bi$_4$Br$_4$ can be identified in α-Bi$_4$I$_4$ at both the $\bar{\Gamma}$ point and the $\bar{Z}$ point (see Fig. 3**b** and 3**c**). α-Bi$_4$I$_4$ also possesses the double-layer structure and the feature of double band inversions after considering SOC (see Supplementary Fig. 16 for details). Thus, such surface states of (100) plane also originate from the double band inversions regardless of their different symmetry-indicator topological invariants.

The first-order topological insulator and second-order topological insulator, which possess the varied bulk-boundary correspondence with 2D surface states and 1D hinge states, respectively, are intuitively regarded to be incompatible in one material due to their order sequence in topological theory and their dimensional discrepancy in real space. Our work providing compelling evidence for the realization of this composite topological phase, the coexistence of HOTI and STI, in a quasi-1D system with the parent phase of HOTI by varying the halide element ratio and the subsequent topological indices assigned at different TRI momenta in BZ. Moreover, a rich phase diagram containing abundant topological phases: high-order topology, weak topology, the unusual dual topology, and trivial/weak topology, is established following the multiple TPTs in this quasi-1D system, which provides an ideal material platform for developing topological materials and quantum devices that could have immense benefits for the fields of topological physics and materials science.

## Methods

1. **Single crystal synthesis and elemental identification**. Solid-state reaction and the chemical vapor transport method were applied to grow the Bi$_4$(Br$_{1-x}$I$_x$)$_4$ single crystals. Highly pure Bi, BiBr$_3$, and BiI$_3$ powders were mixed under Ar atmosphere in a glove box and sealed in a quartz tube under vacuum, where the mole ratio of BiI$_3$ to BiBr$_3$ were adjustable from 0% to 100%. The mixture was placed in a two-heating-zone furnace with the temperature gradient from 558 K to 461 K for 72 h. Monocrystal Bi$_4$(Br$_{1-x}$I$_x$)$_4$ nucleated at the high-temperature side of the quartz after cooling down. The nominal and actual I content of each sample were identified by the mole ratio between the raw materials BiBr$_3$/BiI$_3$ and the energy dispersive spectroscopy (EDS) results for the synthesized single crystals, respectively.

2. **XRD and HAADF-STEM characterizations**. XRD measurements were performed in an AERIS PANalytical X-ray diffractometer at room temperature. The cleaved (001) surfaces were carefully set to be parallel with the sample stage, and the divergence slit was set at 1/8° with beta-filter Ni accompanied by Cu radiation. For STEM characterization, the transmission electron microscopy (TEM) lamella was prepared by using a Zeiss Crossbeam 550 focused-ion-beam (FIB)-scanning electron microscopy (SEM). Then, the characterization was conducted on a probe and image corrected FEI Titan Themis Z microscope equipped with a hot-field emission gun working at 300 kV.

3. **STM measurement.** The STM measurements were carried out using a low-temperature ultrahigh vacuum (UHV)

STM/scanning near-field optical microscopy system (SNOM1400, Unisoku Co.), where the bias voltages were applied to the substrate. The step size of the (001) plane is consistent with monoclinic structure as $c\times\sin107°/2 = 0.96$ nm, and the step size for the (100) plane equals $a/2 = 0.66$ nm. With the projected crystal structure, the atomic-resolved topographies of the crystal structures of the (001) and (100) surfaces are shown in Fig. 1**c** and 1**e**, respectively. The crystal parameters $a$, $b$, and $c$ of $\alpha'$-$Bi_4Br_4$ obtained by STM measurements were 1.33 nm, 0.44 nm, and 2.01 nm, respectively. The differential conductance spectra, $dI/dV$, were acquired by using a standard lock-in technique with a modulation at 973 Hz. All the STM measurements were performed at 77 K.

4. **ARPES characterization.** ARPES measurements were performed on *in-situ*-cleaved thick and homogeneous crystals, attached on the sample holder with torr seal glue with the exposed (001) or (100) surface on the sample holder. The Helium light (with photon energy ~ 21.2 eV) ARPES characterizations were performed at $T = 6$ K at the Photoelectron Spectroscopy Station in the Beijing Synchrotron Radiation Facility using a SCIENTA R4000 analyzer and in our laboratory using a DA30L analyzer. The total energy resolution was better than 15 meV, and the angular resolution was set to ~0.3°, which gave a momentum resolution of ~0.01 $\pi/a$. Laser-based ARPES measurements were performed using DA30L analyzers and vacuum ultraviolet 7 eV lasers at 77 K in Tsinghua University. The overall energy and angle resolutions were set to 3 meV and 0.2°, respectively.

5. **Electric transport measurements.** The standard four-wire method was used to perform the electric transport measurements. The geometry electrodes were fabricated on the freshly cleaved surfaces of $Bi_4(Br_{1-x}I_x)_4$ single crystals, and silver epoxy was applied to connect the samples and electrodes. The results were collected with a physical property measurement system (Quantum Design, Dynacool 9 T) in our home laboratory.

6. **First-principles calculations.** The first-principles calculations were performed using the Vienna *ab initio* simulation package[40] within the projector augmented wave method[41] and the generalized gradient approximation of the Perdew-Burke-Ernzerhof[42] exchange-correlation functional. The energy cut-off of 300 eV was used, and $9 \times 9 \times 3$ and $11 \times 11 \times 4$ $\Gamma$-centered $k$-grid meshes were adopted for structural relaxation and electronic structure calculations, respectively. $11 \times 11 \times 1$ $\Gamma$-centered $k$-grid meshes were used for the $Bi_4Br_4$ monolayer electronic structure. Employing and fixing the experimental lattice symmetries, the crystal structure of $Bi_4(Br_{1-x}I_x)_4$ was fully relaxed with van der Waals correction until the residual forces on each atom were less than 0.01 eV/Å. The converging criterion on total energy is $1 \times 10^{-6}$ eV in all calculations. The SOC effect has been considered in our calculations. A tight-binding (TB) Hamiltonian based on the maximally localized Wannier functions (MLWF)[43] was constructed to further calculate the surface states with the WannierTools package[44]. The $Bi_4(Br_{1-x}I_x)_4$ models were generated by substituting different number of Br with I in a periodic cell containing eight Bi atoms and eight Br atoms.

**Data availability:** All data needed to evaluate the conclusions in the paper are present in the paper. Additional data are available from the corresponding authors upon reasonable request.

# References

1. Kane, C. L. & Mele, E. J. $Z_2$ Topological Order and the Quantum Spin Hall Effect. *Phys. Rev. Lett.* **95**, 146802 (2005).
2. Fu, L., Kane, C. L. & Mele, E. J. Topological Insulators in Three Dimensions. *Phys. Rev. Lett.* **98**, 106803 (2007).
3. Moore, J. E. & Balents, L. Topological invariants of time-reversal-invariant band structures. *Phys. Rev. B* **75**, 121306 (2007).
4. Roy, R. Topological phases and the quantum spin Hall effect in three dimensions. *Phys. Rev. B* **79**, 195322 (2009).
5. Hasan, M. Z. & Kane, C. L. Colloquium: Topological insulators. *Rev. Mod. Phys.* **82**, 3045-3067 (2010).
6. Lv, B. Q., Qian, T. & Ding, H. Experimental perspective on three-dimensional topological semimetals. *Rev. Mod. Phys.* **93**, 025002 (2021).
7. Chen, Y. L. et al. Experimental Realization of a Three-Dimensional Topological Insulator, $Bi_2Te_3$. *Science* **325**, 178-181 (2009).
8. Zhang, H. et al. Topological insulators in $Bi_2Se_3$, $Bi_2Te_3$ and $Sb_2Te_3$ with a single Dirac cone on the surface. *Nat. Phys.* **5**, 438-442 (2009).
9. Bernevig, B. A., Hughes, T. L. & Zhang, S. C. Quantum Spin Hall Effect and Topological Phase Transition in HgTe Quantum Wells. *Science* **314**, 1757-1761 (2006).
10. Brüne, C. et al. Quantum Hall Effect from the Topological Surface States of Strained Bulk HgTe. *Phys. Rev. Lett.* **106**, 126803 (2011).
11. Tang, S. et al. Quantum spin Hall state in monolayer 1T'-$WTe_2$. *Nat. Phys.* **13**, 683-687 (2017).
12. Otrokov, M. M. et al. Prediction and observation of an antiferromagnetic topological insulator. *Nature* **576**, 416-422 (2019).
13. Xu, S. Y. et al. Topological Phase Transition and Texture Inversion in a Tunable Topological Insulator. *Science* **332**, 560-564 (2011).
14. Zhang, P. et al. Observation and control of the weak topological insulator state in $ZrTe_5$. *Nat. Commun.* **12**, 406 (2021).
15. Eremeev, S.V. et al. Atom-specific spin mapping and buried topological states in a homologous series of topological insulators. *Nat. Commun.* **3**, 635 (2012).
16. Okuda, T. et al. Experimental Evidence of Hidden Topological Surface States in $PbBi_4Te_7$. *Phys. Rev. Lett.* **111**, 206803 (2013).
17. Papagno, M. et al. Multiple Coexisting Dirac Surface States in Three-Dimensional Topological Insulator $PbBi_6Te_{10}$. *ACS Nano* **10**, 3518-3524 (2016).
18. Nayak, J. et al. Multiple Dirac cones at the surface of the topological metal LaBi. *Nat. Commun.* **8**, 13942 (2017).
19. He, S., Yang, M. & Wang, R.-Q. Double band-inversions of bilayer phosphorene under strain and their effects on optical absorption. *Chin. Phys. B* **27**, 047303 (2018).
20. Lin, C. et al. Visualization of the strain-induced topological phase transition in a quasi-one-dimensional superconductor $TaSe_3$. *Nat. Mater.* **20**, 1093-1099 (2021).
21. Noguchi, R. et al. Evidence for a higher-order topological insulator in a three-dimensional material built from van der Waals stacking of bismuth-halide chains. *Nat. Mater.* **20**, 473-479 (2021).
22. Ma, X. et al. $Ta_2NiSe_5$: A candidate topological excitonic insulator with multiple band inversions. *Phys. Rev. B* **105**, 035138 (2022).
23. Gupta, S.K., Kore, A., Sen, S.K. & Singh, P. Coexistence of giant Rashba splitting, multiple band inversion, and multiple Dirac surface states in the three-dimensional topological insulator $X$SnBi ($X$=Rb,Cs). *Phys. Rev. B* **107**, 075143 (2023).
24. Chakraborty, A. et al. Observation of highly anisotropic bulk dispersion and spin-polarized topological surface states in $CoTe_2$. *Phys. Rev. B* **107**, 085406 (2023).
25. Shumiya, N. et al. Evidence of a room-temperature quantum spin Hall edge state in a higher-order topological insulator. *Nat. Mater.* **21**, 1111-1115 (2022).
26. Zhong, J. et al. Facet-dependent electronic quantum diffusion in the high-order topological insulator $Bi_4Br_4$. *Phys. Rev. Appl.* **17**, 064017 (2022).
27. Zhao, W. et al. Topological electronic structure and spin texture of quasi-one-dimensional higher-order topological insulator $Bi_4Br_4$. *Nat. Commun.* **14**, 8089 (2023).
28. Zhong, J. et al. Observation of Anomalous Planar Hall Effect Induced by One-Dimensional Weak Antilocalization. *ACS Nano* **18**, 4343–4351 (2024).
29. Dikarev, E. V., Popovkin, B. A. & Shevelkov, A. V. New polymolecular bismuth monohalides. Synthesis and crystal structures

of Bi$_4$Br$_x$I$_{4-x}$ (*x* = 1, 2, or 3). *Russ. Chem. Bull.* **50**, 2304-2309 (2001).

30. Oh, J. S. et al. Ideal weak topological insulator and protected helical saddle points. *Phys. Rev. B* **108**, L201104 (2023).
31. Autes, G. et al. A novel quasi-one-dimensional topological insulator in bismuth iodide *β*-Bi$_4$I$_4$. *Nat. Mater.* **15**, 154-158 (2016).
32. Noguchi, R. et al. A weak topological insulator state in quasi-one-dimensional bismuth iodide. *Nature* **566**, 518-522 (2019).
33. Huang, J. et al. Room-Temperature Topological Phase Transition in Quasi-One-Dimensional Material Bi$_4$I$_4$. *Phys. Rev. X* **11** 031402 (2021).
34. Yang, M. et al. Large-Gap Quantum Spin Hall State and Temperature-Induced Lifshitz Transition in Bi$_4$Br$_4$. *ACS Nano* **16**, 3036-3044 (2022).
35. Zhou, J. J., Feng, W., Liu, C. C., Guan, S. & Yao, Y. Large-gap quantum spin Hall insulator in single layer bismuth monobromide Bi$_4$Br$_4$. *Nano Lett.* **14**, 4767-4771 (2014).
36. Zhong, J. et al. Towards layer-selective quantum spin hall channels in weak topological insulator Bi$_4$Br$_2$I$_2$. *Nat. Commun.* **14**, 4964 (2023).
37. Po, H.C. Symmetry indicators of band topology. *J. Phys.: Condens. Matter* **32**, 263001 (2020).
38. Hossain, M.S. et al. A hybrid topological quantum state in an elemental solid. *Nature* **628**, 527-533 (2024).
39. Bradlyn, B. et al. Topological quantum chemistry. *Nature* **547**, 298-305 (2017).
40. Kresse, G. & Furthmuller, J. Efficient iterative schemes for ab initio total-energy calculations using a plane-wave basis set. *Phys. Rev. B* **54**, 11169 (1996).
41. Blöchl, P. E. Projector augmented-wave method. *Phys. Rev. B* **50**, 17953 (1994).
42. Perdew, J. P., Burke, K. & Ernzerhof, M. Generalized Gradient Approximation Made Simple. *Phys. Rev. Lett.* **77**, 3865 (1996).
43. Souza, I., Marzari, N. & Vanderbilt, D. Maximally localized Wannier functions for entangled energy bands. *Phys. Rev. B* **65**, 035109 (2001).
44. Wu, Q., Zhang, S., Song, H.-F., Troyer, M. & Soluyanov, A. A. WannierTools: An open-source software package for novel topological materials. *Comput. Phys. Commun.* **224**, 405 (2018).

## Acknowledgements

We thank Dr. Lexian Yang for accessing Laser-ARPES system and fruitful discussion in Tsinghua University. We are grateful to the Analysis & Testing Center of Beihang University for the facilities and the scientific and technical assistance. This work was supported by the National Key R&D Program of China (2022YFB3403400 and2018YFE0202700), the Fundamental Research Funds for the Central Universities (Grant Nos. YWF-23SD00-001 and YWF-22-K-101), and the National Natural Science Foundation of China (62141604, 12004321, 12274016, 12004321, and 52473287).

## Author contributions

J.C.Z. and Y.D. conceived the experiments. J.Y.Z., M.Y., W.X.Z., and K.Y.Z., carried out ARPES measurements. J.C.Z., Y.D.L., and D.M. performed the STM/STS measurements. N.Y.C. carried out HAADF-STEM measurements. Z.J.S., J.F.W., X.Z., L.F.Z, and Z.P.H. performed *ab initio* calculations. W.Z. and M.Y. synthesized and characterized the single crystals. J.H.L. and W.C.H. analysed data. J.Y.Z. and M.Y. wrote the first draft of the paper. J.C.Z., Z.P.H., and Y.D. contributed to the revision of the manuscript. J.C.Z. supervised this work. All authors contributed to the scientific planning and discussion.

**Competing interests:** The authors declare no competing interests.

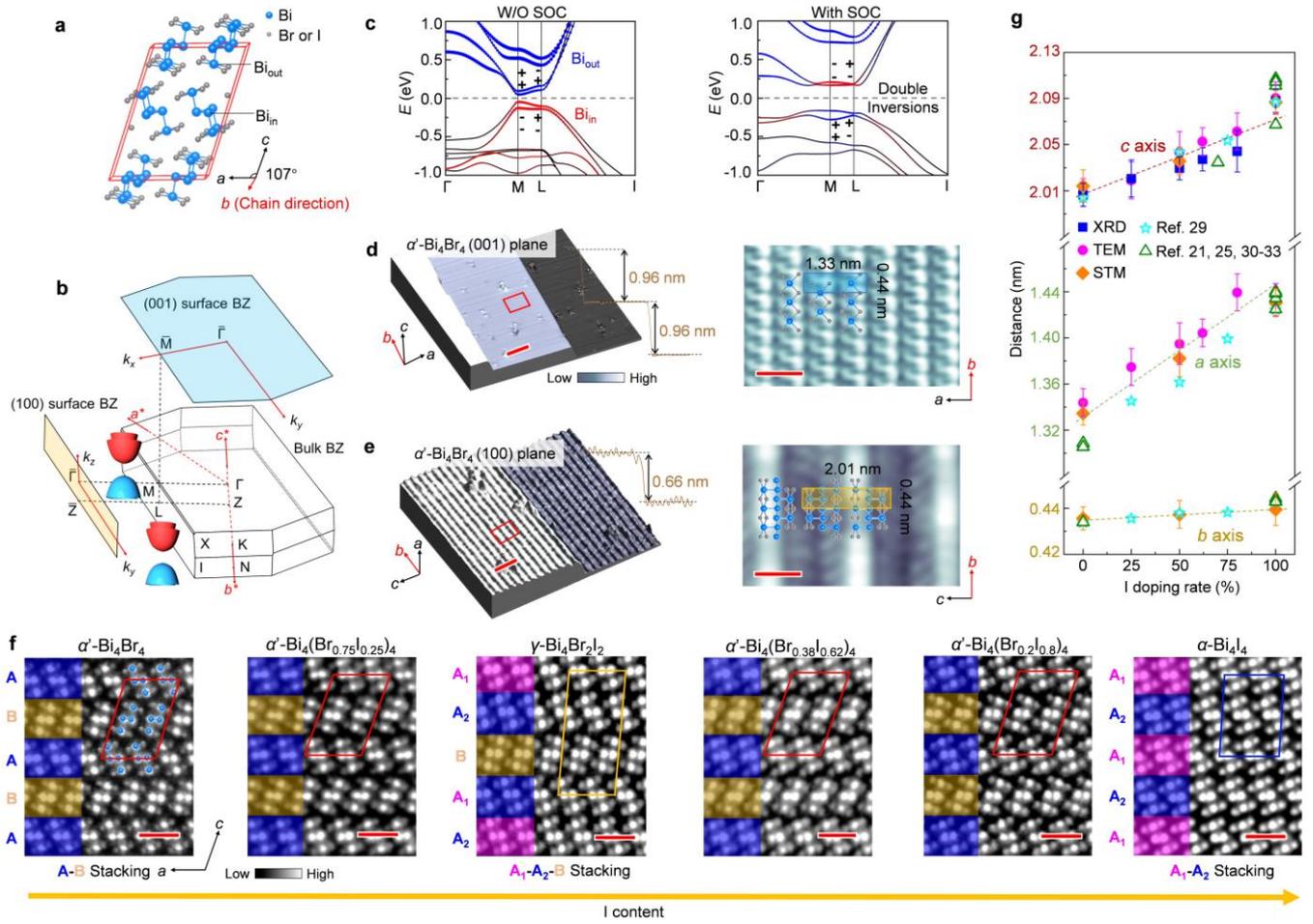

**Fig. 1 | Crystal structures of Bi$_4$(Br$_{1-x}$I$_x$)$_4$ and calculated band structures of Bi$_4$Br$_4$. a**, Crystal structure with the unit cell labelled by red lines of α' phase. The Bi and Br/I atoms are marked by blue and grey balls, respectively. The two sites of Bi atoms are denoted as Bi$_{in}$ and Bi$_{out}$, respectively. The chain direction is along the crystal $b$ axis direction and highlighted in red. **b**, Three-dimensional (3D) Brillouin zone (BZ) with (001) and (100) projected surface BZs. Schematics represent two nondegenerate bands near the Fermi level at $M$ and $L$. **c**, Calculated bulk bands of α'-Bi$_4$Br$_4$ without (W/O) spin-orbit coupling (SOC) (left panel) and with SOC (right panel), respectively. Parity eigenvalues of each band are labelled. The red and blue dots represent the contributions of the $p$-orbitals of Bi$_{in}$ and Bi$_{out}$ atoms, respectively. **d** and **e**, Left: 3D topography of the (001) surface and the (100) surface of α'-Bi$_4$Br$_4$ with step height of 0.96 nm and 0.66 nm, respectively. Scale bar, 5 nm. Right: Atomic-resolved topography of the red squares in the left panels. Scale bar, 1 nm. The blue rectangle and yellow rectangle denote the (001) plane and (100) plane of the unit cell with corresponding atoms, respectively. Lattice constants of $a$, $b$, and $c$ are ~ 1.33 nm, ~ 0.44 nm, and ~ 2.01 nm, respectively. **f**, Atomic-resolved images of the (010) planes of the samples with different I content. Scale bar, 1 nm. The blue, yellow, and pink colours present different layers in each stacking mode. The red, yellow, and blue rhombuses stand for the (010) planes of the unit cells for α' phase, γ phase, and α phase, respectively. **g**, Crystal parameters summarized from our data as well as the previous references [21, 25, 29-33]. Error bars are determined by average variation of X-ray diffraction (XRD), transmission electron microscopy (TEM), and scanning tunnelling microscopy (STM) results, respectively. All the lattice constants increase with I content.

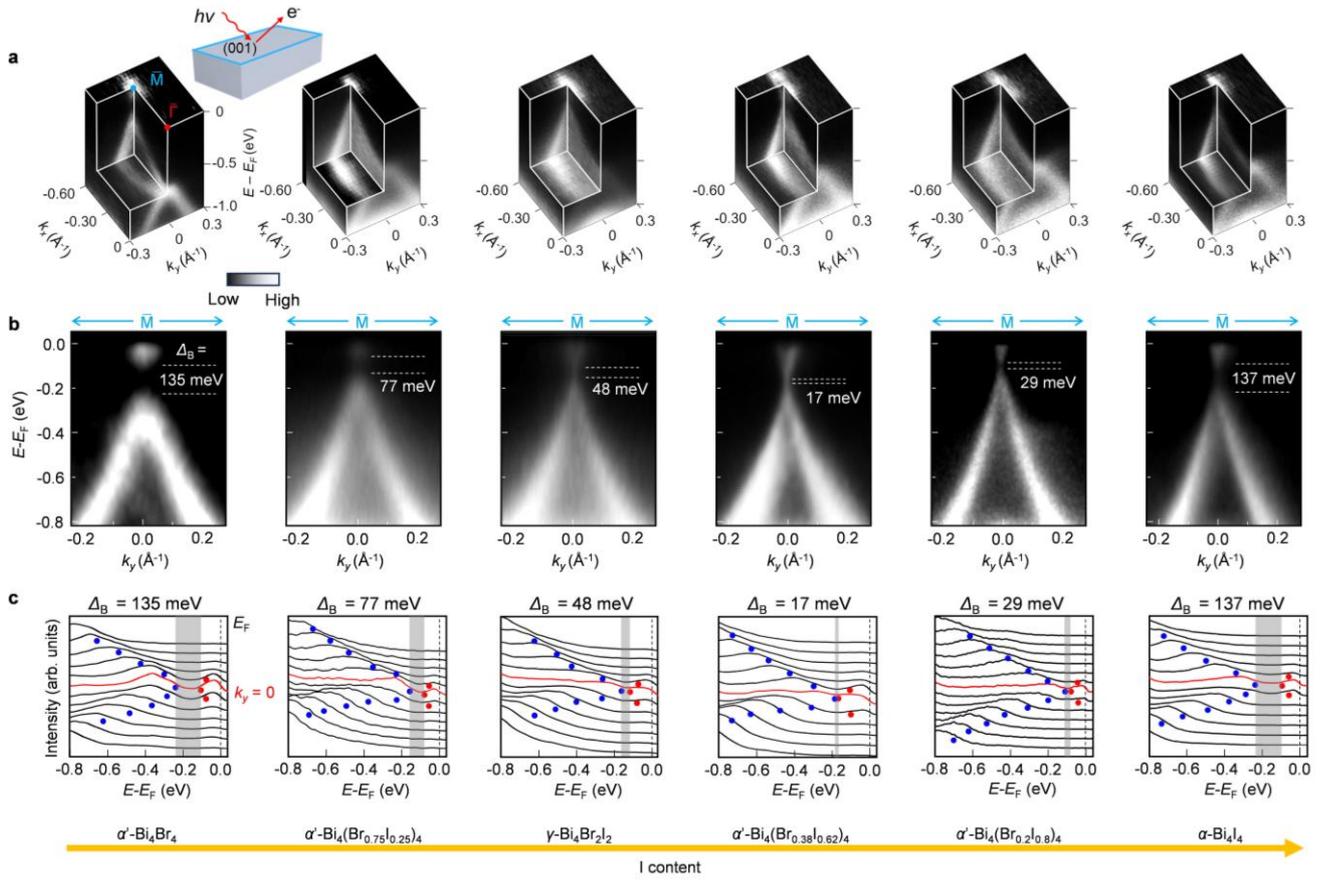

**Fig. 2 | ARPES spectra of the cleaved (001) planes of Bi$_4$(Br$_{1-x}$I$_x$)$_4$ crystals. a,** $E$-$k_x$-$k_y$ dispersion spectra measured at the *in-situ*-cleaved (001) surfaces of Bi$_4$(Br$_{1-x}$I$_x$)$_4$ crystals with $x$ ranging from 0 to 1(from left to right). The (001) projected time-reversal-invariant (TRI) momenta at the $\bar{M}$ point and the $\bar{\Gamma}$ point are labelled by a blue dot and a red dot, respectively. The inset shows a schematic diagram of the experimental setup for the (001) surface. **b,** The $E$-$k_y$ dispersion at $\bar{M}$ along the chain direction. The edge-to-edge gap of bulk bands is indicated by the white dashed lines, where the bulk gap firstly shrinks, and then reopens with a transition point around $x \sim 0.7$. **c,** Energy distribution curves (EDCs) of the corresponding spectra in panel **b**. The EDCs of $k_y = 0$ are marked in red. The blue and red dots are obtained by the peak fitting of EDCs according to the Lorentz-type distribution equation (see Supplementary Note 4 for details), indicating the half-maximum edge locations of the valence band and the conduction band, respectively. The bottom arrow is used to indicate the trend for the I content in different crystals.

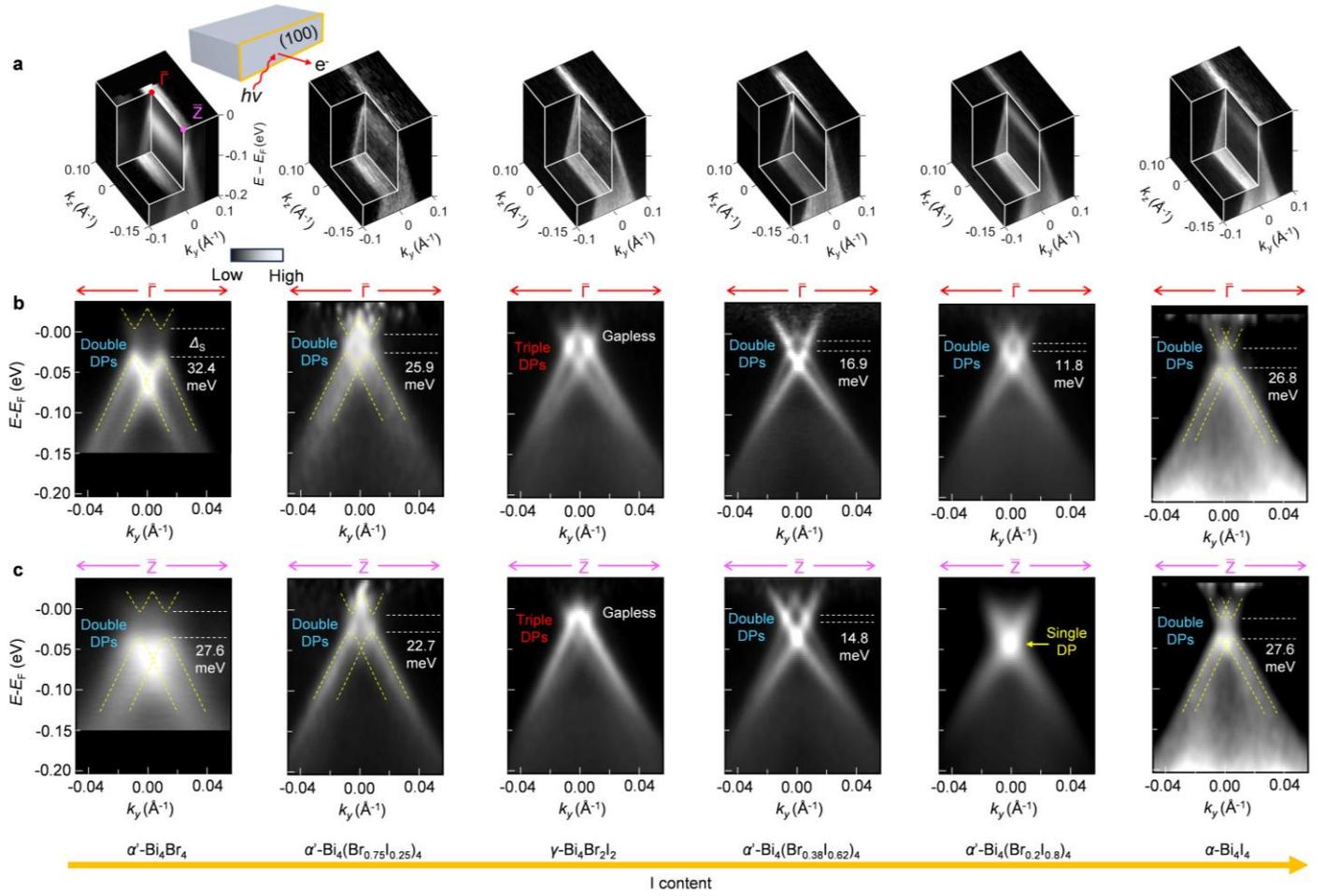

**Fig. 3 | Electronic structures of the (100) planes of $Bi_4(Br_{1-x}I_x)_4$ crystals. a,** $E$-$k_y$-$k_z$ dispersion spectra measured at the *in-situ*-cleaved (100) surfaces of $Bi_4(Br_{1-x}I_x)_4$ crystals with *x* ranging from 0 to 1(from left to right). The inset indicates the experimental setup for the (100) surface characterization. The red dot and pink dot represent the (100) projected $\bar{\Gamma}$ point and $\bar{Z}$ point, respectively. **b** and **c,** High-resolution $E$-$k_y$ dispersions at the $\bar{\Gamma}$ point and $\bar{Z}$ point, respectively. The yellow dashed lines are guides to the eye for the surface state dispersion. The coupling gap $\Delta_S$ is indicated by the white dashed line. There are two nondegenerate Dirac points (DPs) at both the $\bar{\Gamma}$ point and the $\bar{Z}$ points for I concentrations of 0%, 25%, 62%, and 100%. The $Bi_4(B_{0.5}I_{0.5})_4$ crystal exhibits a triple-layer unit cell and is identified as a weak topological insulator (WTI)[36]. The crystal with 80% I content, however, shows double DPs at the $\bar{\Gamma}$ point but a single DP at the $\bar{Z}$ point (labelled by a yellow arrow). The bottom orange arrow denotes the I concentration.

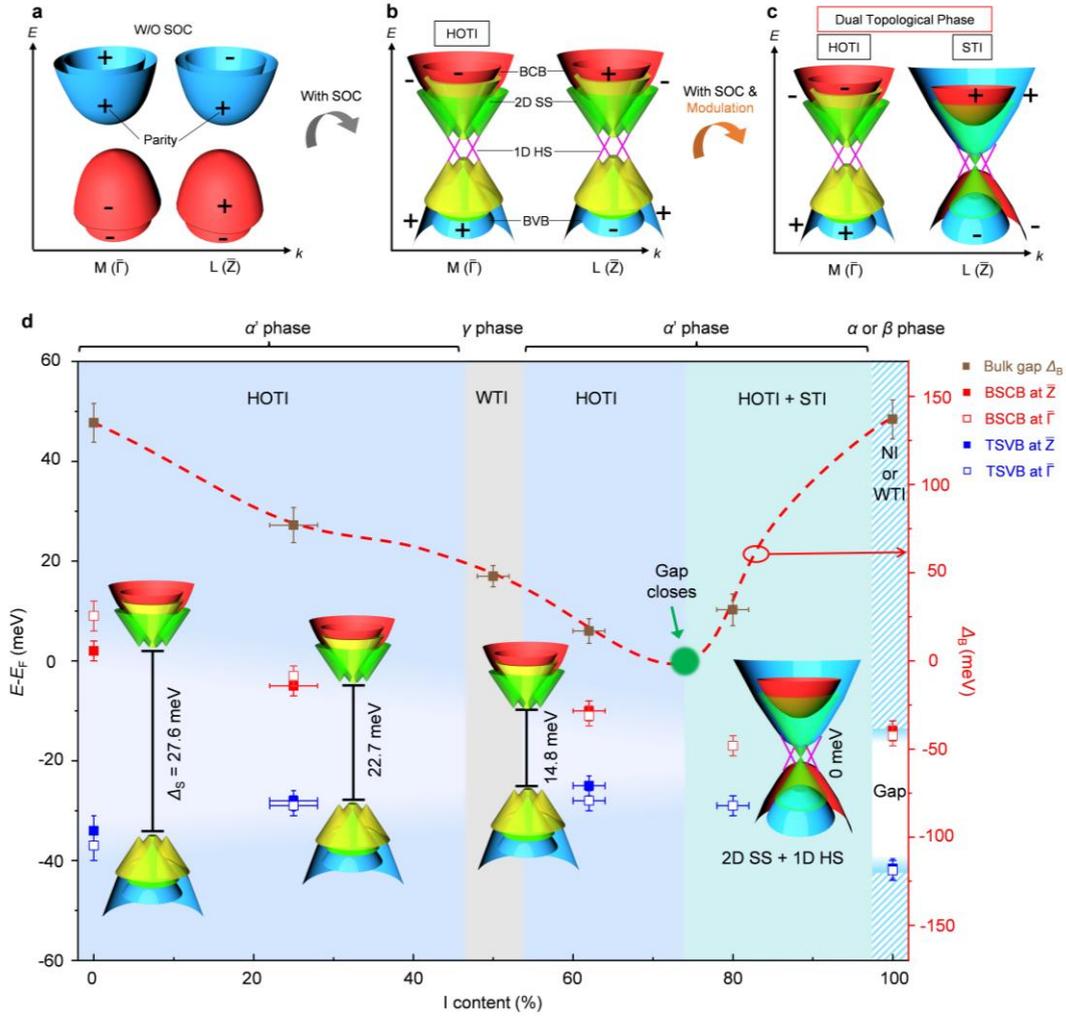

**Fig.4 | Topological phase diagram of the Bi$_4$(Br$_{1-x}$I$_x$)$_4$ system.** Schematic illustration of the band structure of α'-Bi$_4$Br$_4$ at the $M$ ($\bar{\Gamma}$) point and $L$ ($\bar{Z}$) point of the 3D (two-dimensional (2D) (100)) BZ **a,** without SOC and **b,** with SOC. The parity eigenvalues are labelled on each band. The blue bands and red bands are correlated with the bulk structures contributed by the Bi$_{in}$ component and Bi$_{out}$ component, respectively. The green bands and yellow bands stand for the 2D surface states (SS) evoked by the double bulk band inversions at these two TRI momenta. The pink solid lines denote the one-dimensional (1D) hinge states (HS) of the high-order topological insulator (HOTI) state residing in the energy gap induced by the coupling of two nondegenerate SS. The parity eigenvalues switch from (--, ++) to (++, --) at the $M$ point, corresponding to the trivial $\mathbb{Z}_2$ but nontrivial invariant $\mathbb{Z}_4 = 2$. BCB, bulk conduction bands. BVB, bulk valence bands. **c,** Dual topological phase with the coexistence of HOTI and strong topological insulator (STI) states generated by the transition of double band inversions to a single band inversion at the $L$ point of the 3D BZ. **d,** Topological phase diagram of Bi$_4$(Br$_{1-x}$I$_x$)$_4$ and the bulk gap $\varDelta_B$ (brown squares) as a function of I content. The sizes of the SS gap $\varDelta_S$ around $\bar{Z}$ for samples with different I content are also marked. The solid (hollow) red squares and blue squares denote the energy positions of the bottom surface conduction band (BSCB) and top surface conduction band (TSCB) at $\bar{Z}$ ($\bar{\Gamma}$), respectively. NI, normal insulator. Error bars of I content and energy are determined by elemental variation measured by EDS and fitting uncertainty of EDCs, respectively.